\documentclass[%
reprint,
superscriptaddress,
amsmath,amssymb,
aps,
pra,
floatfix,
]{revtex4-1}

\usepackage{graphicx}
\usepackage{xcolor}
\usepackage[normalem]{ulem}
\usepackage{dcolumn}
\usepackage{bm}
\usepackage{mathtools}
\usepackage[thinc]{esdiff}
\usepackage{textcomp}
\usepackage{bbold}
\usepackage{braket}
\usepackage{hyperref}
\usepackage{dsfont}
\usepackage{amsthm}

\usepackage{placeins}
\usepackage[T1]{fontenc}

\begin{document}


\title{Low-Dissipation Data Bus via Coherent Quantum Dynamics}

\author{Dylan Lewis}
 \email{dylan.lewis.19@ucl.ac.uk}
\affiliation{
 Department of Physics and Astronomy, University College London, 
 London WC1E 6BT, United Kingdom
}
\author{João P. Moutinho}
\affiliation{Instituto de Telecomunicações, Physics of Information and Quantum Technologies Group, Lisbon, Portugal}
\affiliation{Instituto Superior Técnico, Universidade de Lisboa, Lisbon, Portugal}

\author{António Costa}
\affiliation{International Iberian Nanotechnology Laboratory, 4715-330 Braga, Portugal}

\author{Yasser Omar}
\affiliation{Instituto Superior Técnico, Universidade de Lisboa, Lisbon, Portugal}
\affiliation{Centro de Física e Engenharia de Materiais Avançados (CeFEMA),
Physics of Information and Quantum Technologies Group, Portugal}
\affiliation{Portuguese Quantum Institute, Lisbon, Portugal}

\author{Sougato Bose}
\affiliation{
 Department of Physics and Astronomy, University College London, 
 London WC1E 6BT, United Kingdom
}


\begin{abstract}
The transfer of information between two physical locations is an essential component of both classical and quantum computing. In quantum computing the transfer of information must be coherent to preserve quantum states and hence the quantum information. We establish a simple protocol for transferring one- and two-electron encoded logical qubits in quantum dot arrays. The theoretical energetic cost of this protocol is calculated---in particular, the cost of freezing and unfreezing tunnelling between quantum dots. Our results are compared with the energetic cost of shuttling qubits in quantum dot arrays and transferring classical information using classical information buses. Only our protocol can manage constant dissipation for any chain length. This protocol could reduce the cooling requirements and constraints on scalable architectures for quantum dot quantum computers.
\end{abstract}

\maketitle

\section{\label{sec:intro}Introduction}
Quantum dot quantum computers encode qubit states using electrons isolated in confined regions by electric fields. Efficiently transferring qubit states in these semiconductor-based architectures is a significant unresolved problem for their scalability. Recent proposals have focused on coherently shuttling the electrons~\cite{fujita_coherent_2017, zwerver_shuttling_2022} and on transfer through multiple quantum dots using engineered tunnel couplings~\cite{qiao_coherent_2020}, which employs theoretical results from work in perfect state transfer~\cite{Bose2003,Christandl2005,Bose2008}. 

A distinct but related question is the energetic cost of using electric currents for the transfer of information in semiconductor-based classical computation. Generating the required potential gradients is a major source of energetic cost. 

Landauer's principle states that the energy dissipated in the form of heat to erase a bit of information is $k_B T \log 2$~\cite{landauer_irreversibility_1961}, suggesting a minimum energetic cost for computation. However, any computation can be performed reversibly~\cite{bennett_logical_1973} and the Landauer limit can in theory be surpassed. Despite further work on computing using reversible logic~\cite{fredkin_conservative_1982}, there remain essentially no practical implementations that are both frictionless and fast. Adiabatic computing, which is slow reversible computing, has been proposed but with significant reductions in performance~\cite{hanninen_design_2015,campos-aguillon_mini-mips_2016}. Quantum computing, using unitary evolutions, is inherently reversible and therefore provides a possible platform for low-energy computation. The coherent manipulation of single electrons for classical computation has recently been proposed, with Moutinho et al.~\cite{moutinho_quantum_2022} considering the energetic advantage of using a quantum dot array with Fredkin gates to implement a full adder, raising the pertinent question of whether a classical computer based on using small quantum systems could provide an energetic advantage for classical computation. The logical states of the qubits are one- and two-electron encodings. Motivated by this, we address the question of low-dissipation quantum buses in quantum dot architectures for quantum and classical data. In this model of a quantum data bus, linear chains of qubits can effectively transfer quantum information due to the natural evolution of an interacting Hamiltonian. In theory, quantum state transfer~\cite{Bose2003,Christandl2005} can coherently transfer information via quantum states without necessarily requiring a voltage.

Currently, computers are many orders of magnitude from the Landauer limit, with the most powerful supercomputers consuming on the order of $\textrm{keV}$ to $\textrm{MeV}$ per bit operation~\cite{moutinho_quantum_2022}. Despite the effort in reducing computational energetic costs, the fundamental limits for electron-based computing suggests that the interconnects---fixed wiring---is the \emph{primary} factor limiting the efficiency of computation, potentially orders of magnitude more costly than computation itself~\cite{zhirnov_minimum_2014}. Here, we address this problem directly by proposing a classical bus using coherent quantum dynamics where the energetic cost does not scale with the length of the wire. We establish a protocol for efficient transfer of an electron using perfect state transfer chains and a simple electron separation protocol. This protocol could be used for quantum computation to alleviate some cooling constraints in scalable quantum computing architectures~\cite{boter_spiderweb_2022}. We find the energetic cost of changing the tunnel coupling between two quantum dots and the minimum energetic cost of implementing the protocol. This is compared to our computed minimum energetic costs for shuttling and for classical data buses. We also make a note on the effect of noise in experimental quantum dot arrays.

\section{\label{sec:}Physical model}
For the quantum dot chains that we consider, the logical qubit is encoded in the charge, rather than spin. The state transfer is a state $|\psi_1(t_0)\rangle$ at time $t_0$, initialised on quantum dot 1 in the chain, being transferred to the last quantum dot in the chain at specific time $T$, $|\psi_N(T)\rangle$, with a high fidelity,  $F = | \langle \psi_1(t_0) |\psi_N(T) \rangle |^2$. 

We set the initial time $t_0 = 0$ and the initial state for transfer to $|\psi(0) \rangle = |\psi_1(0)\rangle \otimes |0\rangle \otimes \dots \otimes |0\rangle$. The protocol, in the simplest case, involves simply turning on interactions for specific time $T$ and then turning off interactions. The state is then at the final site $N$ with high fidelity.

\subsection{\label{sec:hubbard_mode}Hubbard model}
The general model for interacting quantum dots is an extended Hubbard Hamiltonian with both capacitive and tunnel coupling,
\begin{multline}
    \frac{H}{\hbar} = \sum_{i, \sigma}\varepsilon_{i} \hat{n}_{i,\sigma} + \sum_{i, \sigma}
    \Gamma_{i} (c^\dagger_{i,\sigma} c_{i+1, \sigma} + c^\dagger_{i+1, \sigma} c_{i, \sigma}) \\ + \sum_{i, \sigma, \sigma^\prime} V_i \hat{n}_{i,\sigma} \hat{n}_{i+1,\sigma^\prime} + \sum_{i} U_i \hat{n}_{i,\uparrow} \hat{n}_{i,\downarrow},
\end{multline}
where $\varepsilon_i$ is the local field applied to quantum dot $i$, $\Gamma_i$ is the tunnel coupling between quantum dots $i$ and $i+1$, $V_i$ is the capacitive coupling between quantum dots $i$ and $i+1$, $U_i$ is the onsite interaction at site $i$, $c_{i,\sigma}$ and  $c^\dagger_{i,\sigma}$ are respectively the annihilation and creation operators of an electron on quantum dot $i$ with spin $\sigma$. The number operator for electrons of spin $\sigma$ is therefore $\hat{n}_{i,\sigma} = c^\dagger_{i,\sigma} c_{i,\sigma}$. 

\subsection{\label{sec:simplified_mode}Simplified models}
In the transfer protocol, we start with an initial state that contains either one or two electrons depending on the logical encoding used.  Hence, assuming the spins of the electrons do not flip and the number of electrons is constant, significant simplifications to the general Hubbard model can be made. For a single-electron logical qubit, we simply have the tunnel-coupling term and local potential,
\begin{equation}
    \label{eq:single_electron_Hamiltonian_H_1}
    \frac{H_1}{\hbar} = \sum_{i=1}^{N-1} \Gamma_{i,i+1}\left(|i\rangle \langle i+1| + h.c.\right) + \sum_{i=1}^{N-1} \varepsilon_i |i\rangle\langle i|,
\end{equation}
where we have defined a single-electron basis for the quantum dot chain: $|i\rangle$ indicates an electron at quantum dot $i$ with the rest of the dots in the chain empty. The spin of the electron is assumed to be unchanged throughout the protocol. The model is more complex for the two electron encoding, we introduce a two-electron state $|i,j\rangle$, with an up spin electron at site $i$ and a down spin electron at site $j$. There are $N$ sites in the set $S$. The basis can therefore be labelled by $p \in S \times S = \{(i,j)~|~i\in S \textrm{ and } j\in S \}$, giving length $N^2$ and can be constructed as $|p\rangle = |i,j\rangle = c_{i,\uparrow}^\dagger c_{j,\downarrow}^\dagger |0\rangle$. The Hamiltonian matrix elements are thus
\begin{equation}
    H_{p,p^\prime} = \langle 0| c_{j,\downarrow} c_{i,\uparrow}  H c_{i^\prime,\uparrow}^\dagger c_{j^\prime,\downarrow}^\dagger |0\rangle
\end{equation}
The anti-commutation relations of fermions must be considered, $\{c_{i,\sigma}, c_{j,\sigma^\prime} \} = \{c_{i,\sigma}^{\dagger}, c_{j,\sigma^\prime}^{\dagger}\} = 0$ and $\{c_{i,\sigma}, c_{j,\sigma^\prime}^{\dagger} \} = \delta_{i,j}\delta_{\sigma,\sigma^\prime}$. With this careful choice of basis, such that the order of creation operators for the up spin and down spin are not permuted, we find
\begin{multline}    
    \label{eq:two_electron_Hamiltonian_H_2}
    \frac{H_2}{\hbar} = \sum_{i,j=1}^{N-1} \Big[ \Gamma_{i,i+1}\left(|i,j\rangle \langle i+1, j| + h.c.\right) \\ + \Gamma_{j,j+1}\left( |i,j\rangle \langle i, j+1| + h.c. \right) \\ + U \delta_{i,j} |i,j\rangle \langle i,j| \\ + V\left(\delta_{i,j+1} + \delta_{i+1,j}\right)|i,j\rangle \langle i,j| \\
    + (\varepsilon_i + \varepsilon_j) |i,j\rangle\langle i,j|\Big],
\end{multline}
where $\delta_{i,j}$ is the Kronecker delta and we have assumed the onsite interaction, $U$, and capacitive coupling, $V$, are the same for all quantum dots. These Hamiltonians live in significantly smaller Hilbert spaces than the full Hubbard model, which is a space that increases exponentially with number of quantum dots $N$. On the other hand, for $H_1 \in \mathcal{H}_1$ and $H_2 \in \mathcal{H}_2$, we have $\dim(\mathcal{H}_1) \sim N$ and $\dim(\mathcal{H}_2) \sim N^2$, which are both significantly simpler to simulate.

\section{\label{sec:single_electron_state_transfer}State transfer with a single electron}
The single-electron Hamiltonian of Eq.~\eqref{eq:single_electron_Hamiltonian_H_1} is equivalent to the Hamiltonian of the single-excitation subspace dynamics of the XY model---a well-studied model for state transfer~\cite{Bose2008}. We consider schemes that limit the use of $\varepsilon_i$ local fields as it would increase the energetic cost of the protocol. The energetic costs of both this protocol and of a classical information bus, are addressed in Section~\ref{sec:energetic_cost}.

In fact, perfect state transfer can be achieved directly with the XY model in a number of ways that do not require local fields. Engineering the spin-chain tunnel couplings can lead to perfect state transfer. This can be seen by rewriting Eq.~\eqref{eq:single_electron_Hamiltonian_H_1} in matrix form, 
\begin{equation}
    \frac{H_1}{\hbar} = 
    \begin{pmatrix}
        \varepsilon_1 & \Gamma_{1,2} & 0 & \cdots &  &  \\
        \Gamma_{1,2} & \varepsilon_2 & \Gamma_{2,3} & 0 & \ddots & \\
        0 &  \Gamma_{2,3} & \varepsilon_3 & \Gamma_{3,4}& 0 &\ddots  \\
        \vdots & 0 & \Gamma_{3,4} & \varepsilon_4 & \Gamma_{4,5} & \ddots & \\
         & \ddots & \ddots & \ddots & \ddots & \ddots & 
    \end{pmatrix}.
\end{equation}
First, note that the raising and lowering operators of a large spin with $s=(N-1)/2$ act on basis states as $\hat{S}_\pm |s,m \rangle = \sqrt{s(s+1) - m(m\pm 1)} |s,m\pm1\rangle$. For given $s$ and quantum dot $1\leq i \leq N$, we find  $m = i-1 -s$. Rotations of the spin can be induced by $S_x$, which is the generator of rotations about the $x$ axis in $\mathrm{SO}(3)$, giving $R_x(\theta) = e^{-i S_x \theta}$. With the relationships above, we see that $H_1$ is equivalent to $ 2 S_x = S_+ + S_-$ by setting $\varepsilon_i = 0$ for all $i$ and 
\begin{align}
    \Gamma_{i,i+1} &= \sqrt{s(s+1) - m(m + 1)} \\
    &= \sqrt{i(N-i)}.
\end{align}
After a time $T=\pi/2$, which gives unitary evolution $U(\pi/2) = e^{-i H_1 \pi/2\hbar } = e^{-i S_x \pi }$, a rotation of $\pi$ around the $x$ axis has been induced. This takes the initial state $|1\rangle$ to the final state $|N\rangle$. Thus performing perfect state transfer in time $T \sim N$, where the tunnel coupling has been scaled such that the largest coupling is 1.  

We can also use the superexchange where the two end qubits are weakly coupled to a relatively strongly-coupled many-body quantum system~\cite{shi_quantum-state_2005, plenio_high_2005, wojcik_unmodulated_2005, wojcik_multiuser_2007}. In this case, the many-body quantum system is the central quantum dots of the chain. While the fidelity of state transfer is high, the superexchange is very slow: if the coupling between the central quantum dots is such that $\Gamma_{i,i+1} \sim 1$, and the coupling of the first and final quantum dots to the central chain is $\Gamma_{1,2} = \Gamma_{N-1,N} = \epsilon$, where $\epsilon \ll 1$, then the transfer time is $T \sim 1/\epsilon^2$. Slow transfer is undesirable for scalable and fast computational architectures because it would require a slow clock frequency.

\section{\label{sec:two_electron_state_transfer}State transfer with two electrons}
State transfer for two electrons is less straight forward. Although too slow for an architecture proposal, we note that transfer using the superexchange is still possible with two electrons. 

Realising perfect state transfer in the same way as the single electron case, with engineered spin chains replicating $S_x$ for a large spin $s$, is not possible for non-zero $U$ and $V$. All diagonal terms would have to be constant (or zero). In the case of two electrons, we would require
\begin{equation}
    \label{eq:two_electron_pst_requirement}
    U \delta_{i,j} + V(\delta_{i,j+1} + \delta_{i+1,j}) + \varepsilon_i + \varepsilon_j = d,
\end{equation}
for all $i$ and $j$. If we consider $|i-j| > 1$, $\varepsilon_i$ must be the same for all $i$. Thus, $U=V=0$ is required for all diagonal elements to be equal. In this case, we could then use the same tunnel couplings as the single electron case and have two non-interacting electrons that both separately perform perfect state transfer at the same time. If $U \ll \Gamma_\textrm{min}$, where $\Gamma_\textrm{min}$ is the smallest coupling $\Gamma_{1,2}$, we have pretty good (not perfect) state transfer---which, given that this work is also motivated by low-dissipation classical computing, would be useful if it is experimentally feasible. For example, a chain of 16 quantum dots, with $U = \Gamma_\textrm{min}/10$, has a fidelity of state transfer of greater than $0.9$. 

We propose a protocol that first involves separating the electrons and then transferring one electron at a time along an engineered chain with perfect state transfer before recombination. 

\subsection{Two electrons and two quantum dots}
\begin{figure}[h]
    \centering
    \includegraphics[scale=0.56]{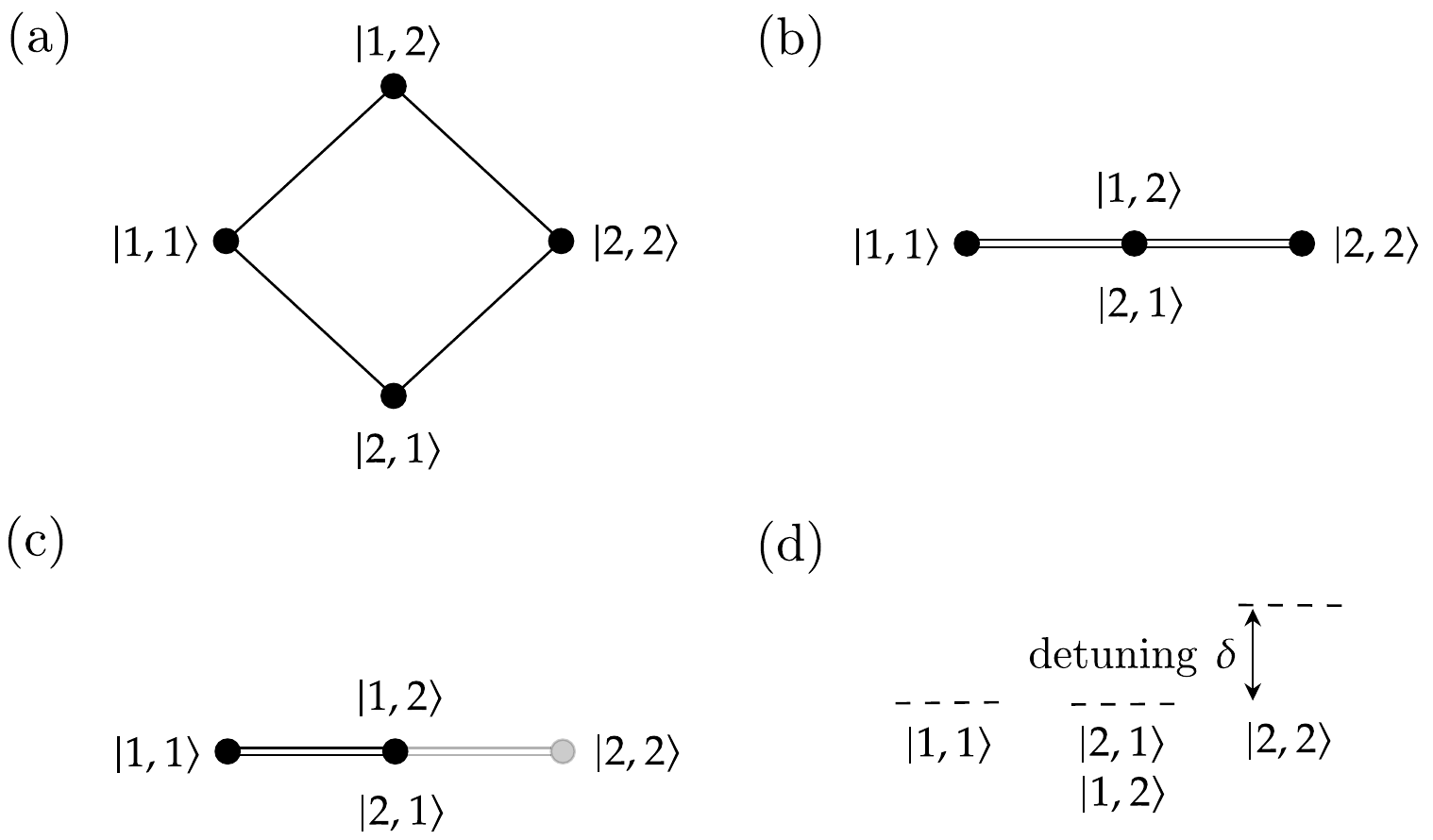}
    \caption{Graphs of states in various representations: (a) all two-electron two quantum dot states are considered where the electrons have opposite spins, (b) simplification of the graph due to symmetry and initial states; (c) state $|2,2\rangle$ is suppressed due to detuning represented in (d). }
    \label{fig:two-electron_two_dots_graph}
\end{figure}

The dynamics of two electrons with two quantum dots can be tuned such that there is high fidelity of electron separation, so one electron on each dot. Perfect state transfer could occur with two electrons on two quantum dots if the Hamiltonian for the evolution of the states---the adjacency matrix of the graph with additional diagonal terms---can be written as 
\begin{equation}
    \frac{H}{\hbar} = \Gamma
    \begin{pmatrix}
        0 & 1 & 1 & 0  \\
        1 & 0 & 0 & 1  \\
        1 & 0 & 0 & 1  \\
        0 & 1 & 1 & 0  \\
    \end{pmatrix} + d\mathds{1},
\end{equation}
where $d$ would have no effect on the evolution, see Fig.~\ref{fig:two-electron_two_dots_graph}(a) for the graph. The analysis can be simplified for certain initial states. The states $|1,2\rangle$ and $|2,1\rangle$ can be considered together because the quantum walk evolution, $U=e^{-iHt/\hbar}$, is symmetric with respect to these states if we start from $|1,1\rangle$ or $|2,2\rangle$, see Fig.~\ref{fig:two-electron_two_dots_graph}(b). This gives the adjacency matrix  
\begin{equation}
    A = \sqrt{2}\Gamma \hbar
    \begin{pmatrix}
        0 & 1 & 0   \\
        1 & 0 & 1  \\
        0 & 1 & 0  \\
    \end{pmatrix},
\end{equation}
which is equal to $2\Gamma S_x$, where $S_x$ is the $x$ spin operator for an $s=1$ boson. Thus, if we assume no detuning between sites, $A$ is equivalent to a rotation around the $x$ axis and, as before, perfect state transfer occurs in time $T = \pi\hbar/2\Gamma$.

If we detune the final state $|2,2\rangle$ from the rest, we suppress the coherent transfer to this site. The dynamics now lead to a high fidelity transfer between $|1,1\rangle$ and a superposition of $|1,2\rangle$ and $|2,1\rangle$, precisely the state required for coherent electron separation. 
To demonstrate the cause of the suppression, consider only the interaction of the superposition of the separated electrons with the $|2,2\rangle$ state, so a two-state system with one of the states detuned by $\delta$. Relabelling the basis states $|0\rangle$ and $|1\rangle$, we have the Hamiltonian
\begin{equation}
    \frac{H}{\hbar} = \Gamma \sigma_x - \frac{\delta}{2} \sigma_z + \frac{\delta}{2} \mathds{1},
\end{equation}
where $\sigma_x$ and $\sigma_z$ are the standard Pauli matrices. We can neglect the identity term as it only adds a global phase. The evolution of the state is therefore
\begin{align}
    U(t) &= e^{-i(\Gamma \sigma_x - \frac{\delta}{2} \sigma_z)t} \\
    &= \cos(n t)\mathds{1} + i \frac{\delta}{2n}\sin(n t) \sigma_z -  i \frac{\Gamma}{n}\sin(n t) \sigma_x,
\end{align}
where $n=\sqrt{\Gamma^2 + (\delta/2)^2}$. When $\delta = m\Gamma$ with $m$ an integer larger than 1, the $\sigma_x$ term is suppressed by $1/\sqrt{1+\frac{m^2}{4}}$. For $m \gg 1$ we have a suppression of $\sim 2/m$ for the rotation term. This leads to a reduction in the fidelity of oscillations from $|0\rangle$ to $|1\rangle$ by approximately $4/m^2$. 
\begin{figure}[t]
    \centering
    \includegraphics[scale=0.42]{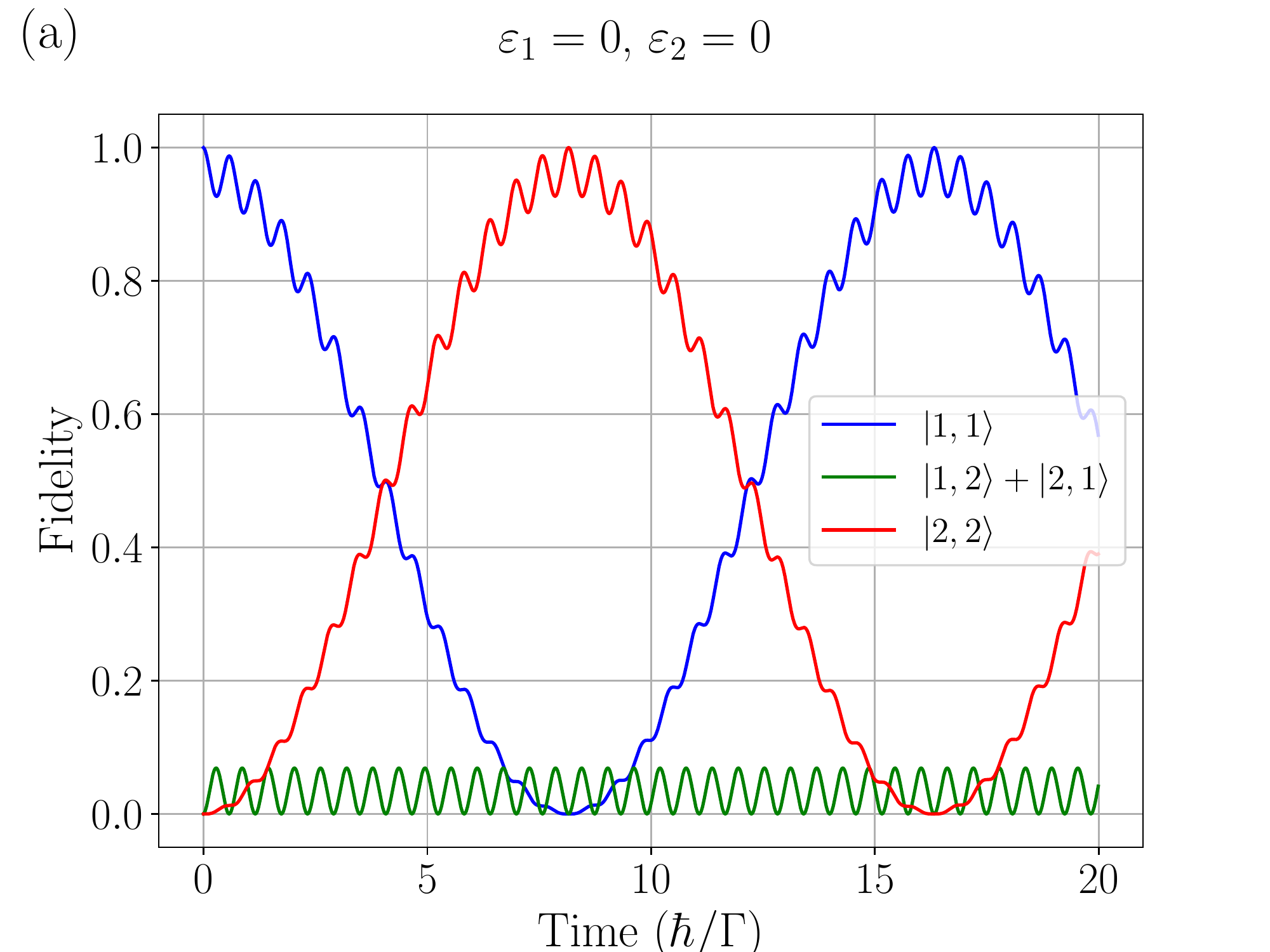}
    \vspace{1em}
    \includegraphics[scale=0.42]{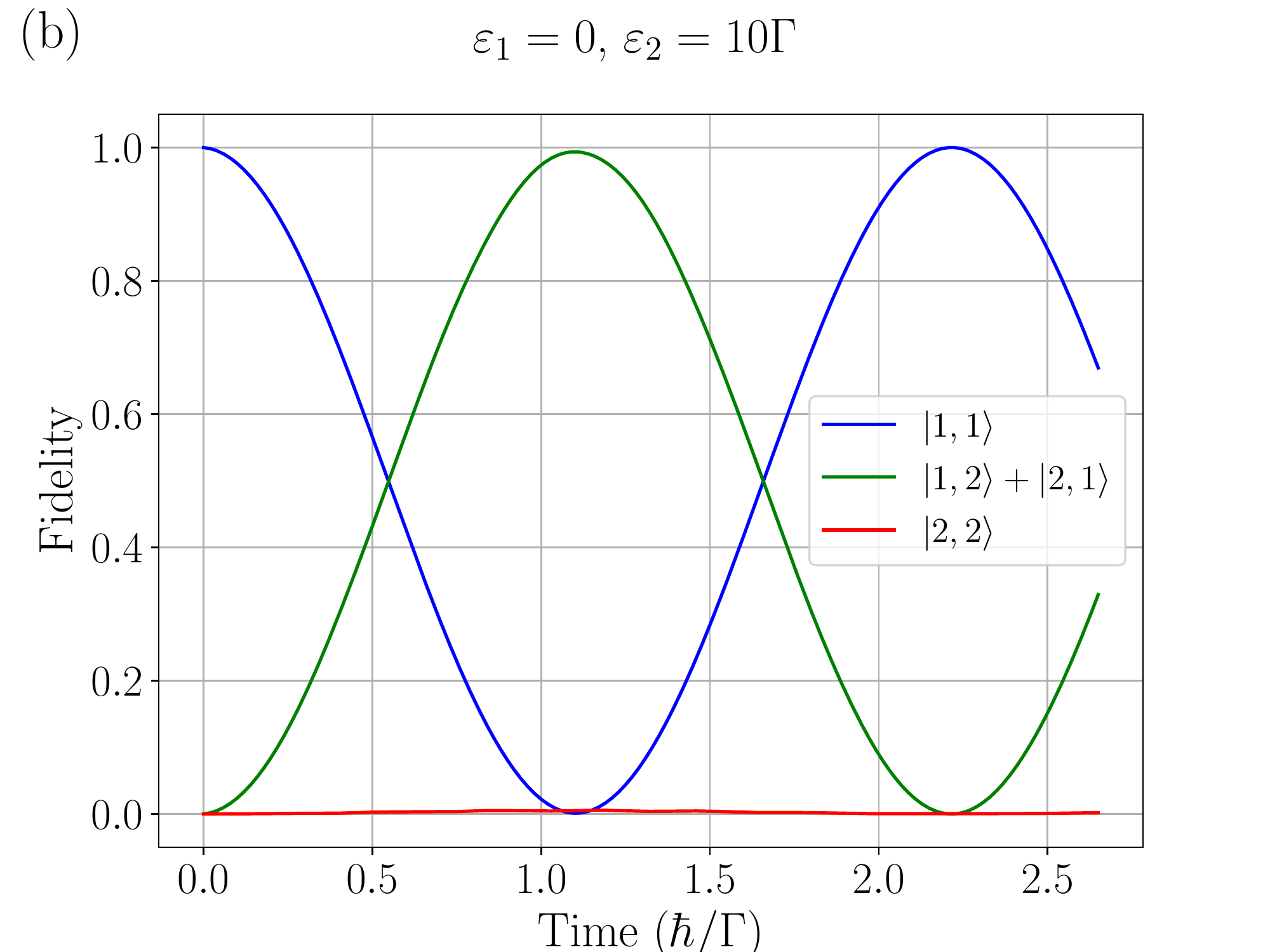}
    \caption{Comparison of the fidelity for the separation of the electrons with $U=20\Gamma$, $V=10\Gamma$ ($\Gamma$ is tunnel coupling) for: (a) no local fields, showing the electrons essentially remain bound together; (b) $\varepsilon_2=10\Gamma$, showing a maximum electron separation fidelity of $0.993$.}
    \label{fig:electron_separation_plots}
\end{figure}
\begin{figure*}[h!t]
    \centering
    \includegraphics[scale=0.8]{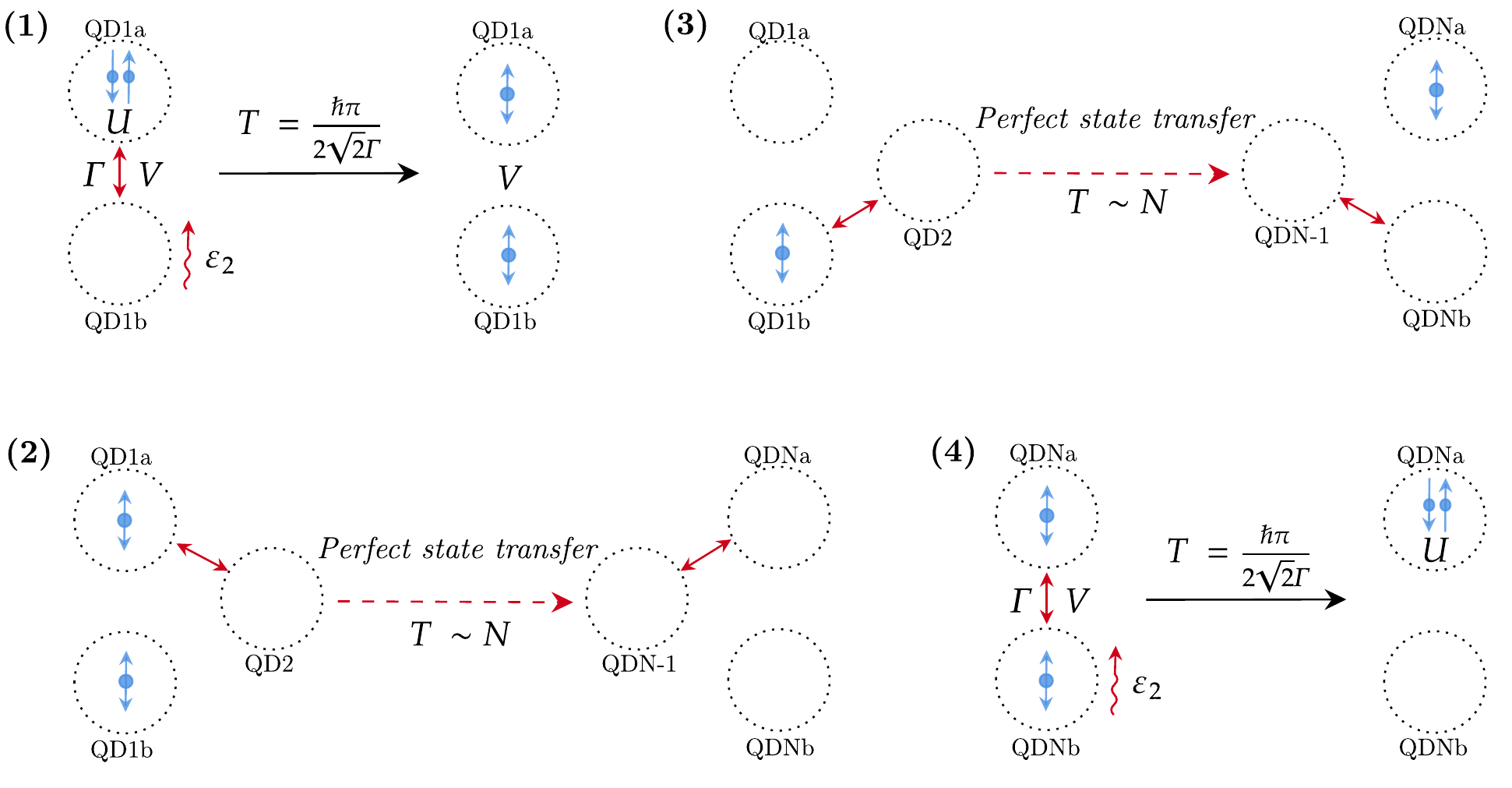}
    \caption{Energy-efficient protocol for transfer of a two-electron logical qubit (or bit): (1) The initial state is loaded into quantum dot 1a (QD1a), and split into an equal superposition of both spins in QD1a and QD1b using electron separation protocol; (2) and (3) Perfect state transfer is used to transfer each electron individually; (4) The electrons are recombined onto QDNa.}    
    \label{fig:protocol_steps}
\end{figure*}
Typical values for electron interaction and capacitive coupling are $U=20\Gamma$ and $V=10\Gamma$, where $\Gamma$ is the tunnel coupling between the two quantum dots~\cite{pedersen_failure_2007, hensgens_quantum_2017}. Applying a local field $\varepsilon_2=10\Gamma$ to only the second quantum dot, detunes the state $|2,2\rangle$ by $\delta=20\Gamma$, while the energy of the states $|1,1\rangle$, $|1,2\rangle$, and $|2,1\rangle$ are all equal. Using the analysis above, we should therefore find a reduction of the fidelity, leaking to the $|2,2\rangle$, state of approximately $0.01$. Numerically, we find a maximum fidelity of separation for the electrons of $0.993$, see Fig.~\ref{fig:electron_separation_plots}. The fidelity can be made higher if we use quantum dots with no capacitive coupling, so $V=0$, and a local field $\varepsilon_2=20\Gamma$, which keeps the energy of the other states equal. The detuning is now $\delta=40\Gamma$, giving an analytical fidelity loss of approximately $0.0025$, which is very close to what we find numerically: a fidelity of separation of $0.998$. The time for the electron separation is that of oscillations in a two state-system with interaction strength $\sqrt{2}\Gamma$---the factor of $\sqrt{2}$ is because there are actually two states $|1,2\rangle$ and $|2,1\rangle$ and therefore two paths between $|1,1\rangle$ to other node of the effective graph. Electron separation therefore occurs in time $T = \pi \hbar / 2\sqrt{2}\Gamma$, which is what we find numerically.

For general $U$ and $V$, to keep the energy of states $|1,1\rangle$, $|1,2\rangle$, and $|2,1\rangle$ equal, we set $\varepsilon_2 = \varepsilon_1 + U - V$, giving the detuning $\delta = 2\varepsilon_2 - 2\varepsilon_1$. The larger $U$ is, while minimising $V$, the larger the difference between $\varepsilon_2$ and $\varepsilon_1$, which increases $\delta$ and therefore the fidelity of electron separation.

Once the electrons have been separated, they are coherently transferred sequentially along the central spin chain with engineered couplings, as in the single-electron case. In theory, this step gives unit fidelity for state transfer. Noise is discussed in Section~\ref{sec:noise}. The electrons are then recombined using the inverse of the separation procedure, with essentially the same fidelity. Fig.~\ref{fig:protocol_steps} shows the steps of the protocol. 

\section{\label{sec:energetic_cost}Energetic cost}
The energetic cost of both the single- and two-electron quantum buses are now considered. As a benchmark, we compare the energetic cost of shuttling electrons, and the lower bound of a data bus in a classical CPU.

With engineered couplings, the transfer of the electron is coherent. Hence the transfer itself does not require an energy source---the reason for an energetic advantage. However, the interactions must be turned on and off, which does have an energetic cost. 

\subsection{Energetic cost of freezing and unfreezing interactions}
We show that the energetic cost of freezing and unfreezing the interactions for a quantum dot system has an optimal energetic cost equivalent to approximately the charging energy of a quantum dot, $E_C$. 
The energetic cost of freezing and unfreezing interactions can be estimated by considering a double quantum dot with two electrons. 

There are two limiting cases: the barrier potential going to zero giving one large quantum dot with two electrons; and the barrier potential being very high giving two isolated quantum dots each with harmonic potentials. In our quantum dot model, we consider the latter case, with a significant barrier. This assumption is reasonable since the preceding protocol for state transfer is in the regime $U \gg \Gamma$. 

\begin{figure}[t]
    \centering
    \includegraphics[scale=0.66]{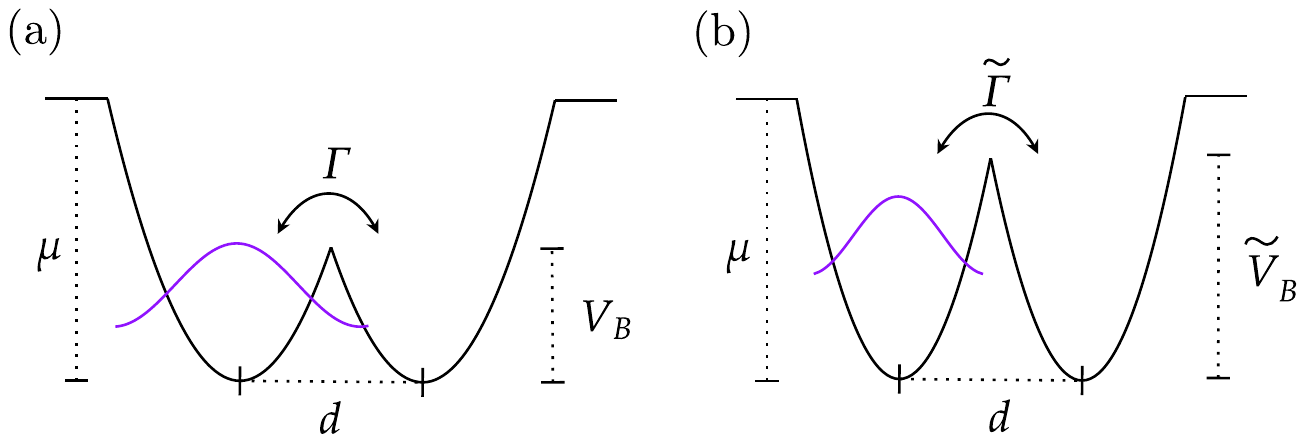}
    \caption{One-dimensional cross section through the centre of the confining harmonic potentials that model a double quantum dot. The confinement is in two dimensions, with radial symmetry around the centre of each of the dots. The size of $V_0$ has been changed between (a) and (b), with (b) showing increased confinement and therefore a higher barrier.}
    \label{fig:biquadratic_potentials}
\end{figure}

Electrons in a double quantum dot can be modelled as a biquadratic potential~\cite{wensauer_laterally_2000, helle_two-electron_2005, pedersen_failure_2007, li_exchange_2010, yang_generic_2011}, see Fig.~\ref{fig:biquadratic_potentials}. The Hamiltonian for two electrons is 
\begin{equation}
    \label{eq:hamiltonian_2_electron_2_dots}
    H = \sum_{j=1}^{2} \left[ \frac{\bm{p_j}^2}{2m^*} + V(\bm{r_j})\right] + \frac{e^2}{4\pi \varepsilon_0 \varepsilon_r} \frac{1}{|\bm{r_1} - \bm{r_2}|},
\end{equation}
where $\bm{p_j}$ and $\bm{r_j}$ are the momentum and position vectors of electron $j$ in two dimensions, $m^*$ is the effective mass, and $V(\bm{r})$ is the potential
\begin{equation}
    V(\bm{r}) = V_0 \min\left[(\bm{r}-\bm{l})^2, (\bm{r}+\bm{l})^2, \mu \right],
\end{equation}
where $\mu$ is the chemical potential, and the dots are located at $\pm \bm{l}$, a distance of $d = 2 l$ apart. 
For large $\mu$, we consider the simplification $V_S(\bm{r}) = \lim_{\mu \rightarrow \infty} V(\bm{r})$.
Fig.~\ref{fig:biquadratic_potentials} shows the form of the potential, with barrier height $V_B = V_0 l^2$ between the dots. Increasing the distance between the harmonic potentials increases the barrier height. Rather than increasing the distance, however, $d$ is fixed and the strength of the harmonic potentials $V_0$ is increased. In the low temperature limit, $k_B T \ll \hbar \omega_0$, the electrons are assumed to be in their ground state. For a two-dimensional harmonic trap with $V_0 = m^* \omega_0^2/2$, the ground state energy, with lowest orbital momentum in the confinement, is $\hbar \omega_0$. We define a dimensionless parameter $\eta = m^* \omega_0 l^2/\hbar$, which is the ratio of the barrier height and half the ground state energy of an electron in a harmonic trap. In this analysis, only $\eta > 1$ is considered, following from $U \gg \Gamma$. In this regime, the Heitler-London (HL) approximation is valid~\cite{yang_generic_2011} since the quantum dots are sufficiently separated. We can thus build the two-electron ground state from single-electron harmonic ground states of model Hamiltonians of the form $h^{(0)}_{L/R} = \bm{p}_1^2/2m^* + m^* \omega_0^2 (\bm{r}_1\pm\bm{l})^2 / 2$. 

If there are no external fields, the ground state of two electrons will always be the singlet state because the spatial wave function is symmetric, i.e. exchange coupling $J \equiv E_T - E_S > 0$~\cite{li_exchange_2010}, where $E_{T/S}$ are the triplet and singlet energies. Further to this, we will consider two electrons on the same quantum dot to calculate the charging energy $E_C$. The lowest energy is when both electrons can occupy the same lowest energy orbital, giving a symmetric spatial wave function, and therefore a singlet spin state. As we are not concerned with the exchange coupling, we only consider the singlet state in the following HL approximation and further analysis.

The ground state of the harmonic potentials for the left and right quantum dots, with Hamiltonians $h^{(0)}_{L/R}$, are 
\begin{equation}
    \varphi_{L/R}(\bm{r}) = \langle \bm{r} | L/R \rangle = \frac{1}{a\sqrt{\pi}} e^{-\frac{1}{2a^2} \left[ (x \pm l)^2 + y^2 \right]},
\end{equation}
where we have defined a Bohr radius $a = \sqrt{\hbar / m^* \omega_0}$, $\bm{r} = (x,y)$, and the dots are centred along the $x$ axis. Using the HL approximation, the ground state of the two-electron double quantum dot is a symmetric spatial superposition of the electrons on different dots.
We define the overlap between the adjacent harmonic potentials, $s = \langle L|R\rangle = e^{- l^2/a^2} = e^{- \eta}$. 
As in Ref.~\cite{burkard_coupled_1999}, the Hund-Mulliken (HM) approximation is used to further include the states with two electrons on the same quantum dot, the (2,0) and (0,2) states, which must also be singlet states in the ground state. The left and right basis states are rotated such that they are orthogonal, $\langle \Phi_L| \Phi_R \rangle = 0$, giving $|\Phi_{L/R}\rangle = \left(|L/R\rangle - g |R/L\rangle \right)/\mathcal{N}$ where $\mathcal{N} = \sqrt{1-2sg+g^2}$ and $g = (1-\sqrt{1-s^2})/s$, such that both $\mathcal{N}$ and $g$ are functions of $\eta$. In this basis, with $\Phi_{L/R}(\bm{r}) = \langle \bm{r} | \Phi_{L/R}\rangle$, the three relevant spatial wave functions are 
\begin{equation}
    \Psi^\textrm{d}_{L}(\bm{r}_1,\bm{r}_2) = \Phi_{L}(\bm{r}_1)\Phi_{L}(\bm{r}_2),
\end{equation}
\begin{equation}
    \Psi^\textrm{d}_{R}(\bm{r}_1,\bm{r}_2) = \Phi_{R}(\bm{r}_1)\Phi_{R}(\bm{r}_2),
\end{equation}
\begin{equation}
    \Psi^\textrm{s}_0(\bm{r}_1,\bm{r}_2) = \frac{1}{\sqrt{2}}\big[ \Phi_{L}(\bm{r}_1)\Phi_{R}(\bm{r}_2) + \Phi_{R}(\bm{r}_1)\Phi_{L}(\bm{r}_2)\big],
\end{equation}
where $\Psi^\textrm{d}_{L/R}(\bm{r}_1,\bm{r}_2)$ indicate the doubly occupied states (2,0) and (0,2), and  $\Psi^\textrm{s}_0(\bm{r}_1,\bm{r}_2)$ indicates both sites being singly occupied (1,1). All these states are symmetric since the states are spin singlets. 

The Hamiltonian is separable for the non-Coulomb terms: $\hat{H} = \hat{h} \otimes \mathds{1} + \mathds{1} \otimes \hat{h} + \hat{C} $, where $\hat{h} = \bm{p}^2/2m^* + V_S(\bm{r})$, and $\hat{C} = e^2 / 4 \pi \varepsilon_r \varepsilon_0 |\bm{r}_1 - \bm{r}_2|$. The tunnelling terms, from the states $\Psi^\textrm{d}_{L}(\bm{r}_1,\bm{r}_2)$ or $\Psi^\textrm{d}_{R}(\bm{r}_1,\bm{r}_2)$ to $\Psi^\textrm{s}_0(\bm{r}_1,\bm{r}_2)$, are then given by the matrix element
\begin{align}
      \hbar \Gamma^{(2)} &= \sqrt{2} \langle \Phi_{L/R} | \hat{h} | \Phi_{R/L}\rangle + \langle \Psi^d_{L/R}| \hat{C} |\Psi^s_0\rangle,
\end{align}
where we have defined a two-electron tunnelling rate, $\Gamma^{(2)}$, including the Coulomb repulsion. If there is no Coulomb repulsion and we ignore the presence of the second electron, we have the `bare' tunnelling rate 
\begin{align}
    \hbar \Gamma &=  \langle \Phi_{L/R} | \hat{h} | \Phi_{R/L}\rangle,\\
    &= \frac{1}{\mathcal{N}^2}\left[(1+g^2) w -2g u\right],
\end{align}
where we have used $w = \langle L | \hat{h} | R\rangle =  \langle R | \hat{h} | L\rangle$ and $ u =\langle L | \hat{h} | L\rangle =  \langle R | \hat{h} | R\rangle$. Furthermore, we find
\begin{equation}
    w = \left(1-\sqrt{\frac{\eta}{\pi}} \right)e^{-\eta} \hbar \omega_0
\end{equation}
and,
\begin{equation}
    u = \left(1- \sqrt{\frac{\eta}{\pi}} e^{-\eta} + \eta~\mathrm{erfc}(\sqrt{\eta})  \right) \hbar\omega_0,
\end{equation}
where $\mathrm{erfc}(\sqrt{\eta})$ is the complementary error function. 

The charging energy, $E_C$, is approximately the difference in energy between having two electrons in the lowest energy level of a single quantum dot and having only one electron in the dot,
\begin{equation}
    E_C \approx E^{(2)}_0 - E^{(1)}_0, 
\end{equation}
where $E^{(1)}_0$ ($E^{(2)}_0$) is the ground state energy of $1$ ($2$) electrons in a single harmonic potential. For a single electron in a harmonic trap, as above, $E_0^{(1)} = \hbar \omega_0$. Two electrons in a single harmonic potential is more complex since the Coulomb repulsion of the two electrons must be considered, and in the double quantum dot model above, we have
\begin{align}
    E^{(2)}_0 &= 2\langle \Phi_{L/R} | \hat{h} | \Phi_{L/R}\rangle + \langle \Psi_{L/R}^{d} | \hat{C} | \Psi_{L/R}^{d} \rangle \\
    &= \frac{2}{\mathcal{N}^2}\left[(1+g^2)u - 2g w \right] + U
\end{align}
for two electrons on either the left or right quantum dot---these are equivalent. The second term in this model is the onsite interaction in the Hubbard model, $U$. For well separated quantum dots, $\eta \gtrsim 1.5$, leading to $u \gg w$ and $g \ll 1$, hence $(1+g^2) \gg 2g$. For the purposes of the following approximations, it is therefore sufficient to give onsite energy due to the momentum and potential as $\langle \Phi_{L/R} | \hat{h} | \Phi_{L/R}\rangle \approx \hbar \omega_0$ and the onsite Coulomb repulsion as $U \approx U_0$, where 
\begin{multline}
    \hbar U_0 = \int d\bm{r}_1 \int d\bm{r}_2 \Big[ \varphi_{L/R}(\bm{r}_1) \varphi_{L/R}(\bm{r}_2) \\ \times C(\bm{r}_1, \bm{r}_2)\varphi_{L/R}(\bm{r}_1) \varphi_{L/R}(\bm{r}_2) \Big],
\end{multline}
with $C(\bm{r}_1, \bm{r}_2) = e^2 / 4 \pi \varepsilon_r \varepsilon_0 |\bm{r}_1 - \bm{r}_2|$. The identity $\frac{1}{|\bm{r}_1 - \bm{r}_2|} = \frac{2}{\sqrt{\pi}}\int_0^{\infty} dt\exp\left\{-t^2 (\bm{r}_1 - \bm{r}_2)^2 \right\}$~\cite{singer_use_1960}
can be used to compute $U_0$, giving $E^{(2)}_0 \approx 2 \hbar \omega_0 +  c \hbar \omega_0$, where $c = (e^2/4\pi\varepsilon_r \varepsilon_0 \tilde{a})/\hbar\omega_0$ and $\tilde{a} = \sqrt{2/\pi}a$; $c$ is the ratio of the Coulomb energy ($e^2/4\pi\varepsilon_r \varepsilon_0 \tilde{a}$) to the confinement energy ($\hbar \omega_0$). Overall, the charging energy is therefore $E_C \approx  (1 +c) \hbar \omega_0$. 

Both $\eta$ and $c$ are dependent on the confinement frequency $\omega_0$, as $\eta = \eta_0 \omega_0$ and $c = c_0/\sqrt{\omega_0}$, where we have introduced the parameters $\eta_0 = l^2 m^* / \hbar$ and $c_0 = \sqrt{\pi m^* /2}e^2/4\pi\varepsilon_r \varepsilon_0\hbar^{3/2}$. After increasing the confinement potential of a single electron by the charging energy, we have the new ground state frequency $\tilde{\omega}_0 = \omega_0 + E_C/\hbar = (2 + c_0/\sqrt{\omega_0})\omega_0$, leading to a change in the ratio of barrier height to half ground state energy, $\tilde{\eta} =  (2+c_0/\sqrt{\omega_0})\eta$. The change in barrier height is therefore dependent on the initial ground state frequency. Typical parameters for GaAs quantum dots are $m^* =0.067 m_e$, $\varepsilon_r = 12.9$, and $\hbar\omega_0=3~\textrm{meV}$~\cite{reimann_electronic_2002}, giving $c_0 = 5.11 \times 10^6~\textrm{Hz}^{\frac{1}{2}}$ and $c = c_0/\sqrt{\omega_0} = 2.39$, hence $\tilde{\omega}_0 \approx 4.39\omega_0$ and $\tilde{\eta} \approx 4.39\eta$. Thus, the energetic cost of charging is $E_C \approx 10~\textrm{meV}$.

The parameter regime is such that the onsite interaction is approximately $U \approx 20 \Gamma$ and since $U \approx U_0 = c \omega_0$, we find $\hbar U \approx 7.2~\textrm{meV}$ and $\hbar \Gamma \approx 360~\textrm{µeV}$, which are both plausible experimental values~\cite{yang_generic_2011, hensgens_quantum_2017}.

\begin{figure}[h]
    \centering
    \includegraphics[scale=0.64]{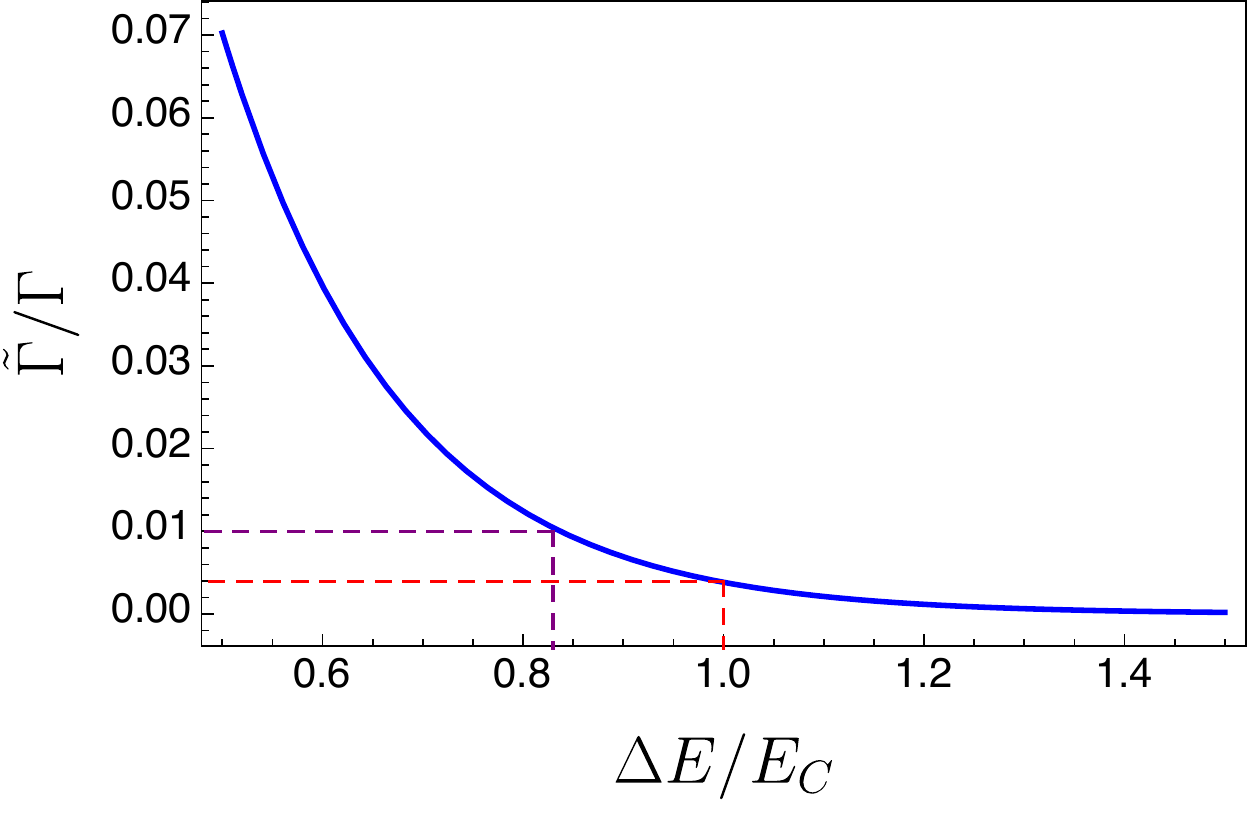}
    \caption{The ratio of the frozen and unfrozen tunnel couplings against the energy applied, $\Delta E$, to increase the confinement potential (the energy is given in units of the charging energy $E_C$). The purple dashed line shows that 1\% of unfrozen tunnel coupling is achieved for $\Delta E= 0.83 E_C$, and the red dashed line shows that 0.4\% of unfrozen tunnel coupling for $\Delta E = E_C$.}
    \label{fig:tunnel_coupling_decrease}
\end{figure}

By enforcing $U = 20\Gamma$, numerically we find $\eta = 1.86$ gives the correct ratio of onsite interaction and tunnel coupling, and therefore $\tilde{\eta} = 8.17$. The new tunnel coupling is $\tilde{\Gamma} \approx 0.0039\Gamma$. Thereby effectively freezing the electron hopping as the tunnelling of a single electron would take approximately 250 times as long. If we define freezing the interactions as approximately reducing the tunnel coupling to $1\%$ of having the interactions unfrozen, we would only require an increase of the confinement energy of approximately $0.83 E_C$, see Fig.~\ref{fig:tunnel_coupling_decrease}. 

The preceding applies in the case that there are two electrons. When there is only one electron the charging energy is significantly less, as we take $c \rightarrow 0$, giving $E_C^{(1)} = \hbar \omega_0$, $\tilde{\omega}_0 = 2 \omega_0$, and $\tilde{\eta} = 2 \eta$. Assuming the same tunnelling strength of $\hbar \Gamma = 360~\textrm{µeV}$ again leads to $\eta = 1.86$ and therefore the new tunnelling is $\tilde{\Gamma} \approx 0.22 \Gamma $, which is much greater than our definition of freezing the tunnelling. In fact, in order to reach the equivalent reduction in tunnelling as in the two-electron case, we must increase the confinement by about $4.5 E_C \approx 13.5~\textrm{meV}$. Additional energy is required because with only one electron the confinement potentials are less giving a lower central barrier due to the constant distance between the dots---see Fig.~\ref{fig:biquadratic_potentials}.  

\subsection{Shuttling}
Shuttling is a proposal for transporting electrons in semiconductor devices for scalable quantum computation~\cite{boter_spiderweb_2022}. An early proposal involves the electron being shuttled by a surface acoustic wave~\cite{hermelin_electrons_2011}. Subsequent proposals have generally used arrays of quantum dots with tunable metal barrier gates to lower and raise the tunnelling rate between neighbouring dots, inducing a transfer of the electron through the dots sequentially~\cite{baart_single-spin_2016, fujita_coherent_2017, mills_shuttling_2019, buonacorsi_simulated_2020, ginzel_spin_2020, seidler_conveyor-mode_2021}. 

For a fair comparison of the energetics, we assume the quantum dots and spacing between them for shuttling are the same as our state transfer protocol. The energetic cost of shuttling can therefore be considered the sequential loading and unloading of the quantum dots to coherently move the electrons along the chain. Although in practice this may be achieved with a separate barrier potential and raising and lowering the chemical potential of the quantum dots, in the best case, this would be energetically equivalent to freezing and unfreezing the tunnelling between adjacent quantum dots. The energetic cost of shuttling is therefore at least $E_\textrm{shuttling}^{(2)} = 2 E_C N \approx 20 N~\textrm{meV}$ for the two-electron encoding, and $E_\textrm{shuttling}^{(1)} = E_C N \approx 13.5 N~\textrm{meV}$ for the single-electron encoding.

\subsection{Perfect state transfer scheme}
The full energetic cost of our scheme includes freezing and unfreezing interactions, but also the cost of applying the local potential $\varepsilon_2$ for separation and recombination of the electrons (see Fig.~\ref{fig:protocol_steps}, steps 1 and 4). The local potentials applied are $\delta \gg \Gamma$, thus as a worst case estimate would change the ground state energies of the electrons by $E_\delta \approx \hbar \delta$. The total energetic cost of our protocol for two-electron encoding is therefore $E_\textrm{PST}^{(2)} = 4 E_C + 2 E_\delta$, independent of the length of the quantum dot chain $N$ (more accurately, $N+2$ due to the additional quantum dot required at each end of the chain). In the worst case, $\delta = 40\Gamma$ and therefore $E_\textrm{PST}^{(2)}\approx 108~\textrm{meV}$ for the two-electron logical qubit encoding.

A single electron logical qubit encoding would only require the energetic cost of a single step 2 or 3 from Fig.~\ref{fig:protocol_steps}. Therefore, we find an energetic cost of $E_\textrm{PST}^{(1)}\approx 54~\textrm{meV}$, half that of the two-electron logical encoding. 
\begin{figure}[h]
    \centering
    \includegraphics[scale=0.48]{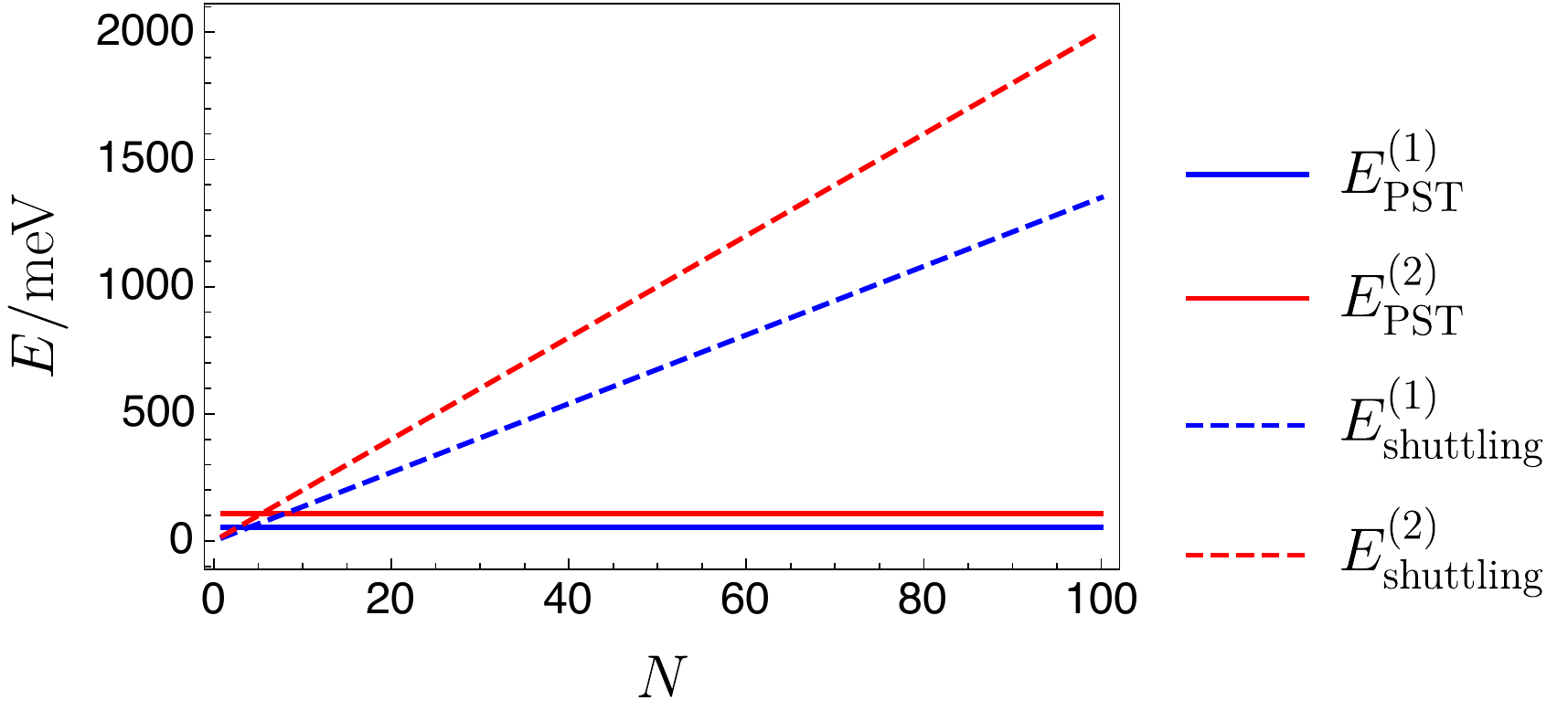}
    \caption{Comparison of the theoretical energetic costs for the protocol with perfect state transfer (PST) and the shuttling method for logical information with both one and two electron encodings.}
    \label{fig:energetics_comparison}
\end{figure}

\subsection{\label{sec:classical_wire_energetics}Lower bound for classical wire}
A lower bound for an interconnect in a CPU can be estimated by the energy required to charge the metal wire, $CV^2$, where we treat the wire as a capacitor with capacitance $C \sim \varepsilon_0 L$, with vacuum permittivity  $\varepsilon_0$ and $L$ is the length of the wire. The minimal distinguishable voltage is $V\sim k_B T/e$~\cite{zhirnov_minimum_2014}, where $k_B$ is the Boltzmann constant and $T$ is temperature, which we assume to be room temperature because in the cold regime of the quantum dots, we would have to consider the quantum effects of the wire. In the cold regime, we have investigated shuttling instead. The size of quantum dots with 3~meV is approximately 100~nm~\cite{reimann_electronic_2002}. Hence, we find the lower bound of $E_\textrm{classical} > 3.7 N~\textrm{meV}$, where $N$ is the equivalent number of $100~\textrm{nm}$ quantum dots for the interconnect. This bound is of course very conservative and in reality far more energy is required in current CPUs, as discussed in the introduction. However, it is already of the same order as quantum coherent buses and, crucially, it scales with $N$, thus showing the advantage of the perfect state transfer protocol.

\section{\label{sec:noise}Noise}
We have established this advantage in the case that there is no noise. There are several sources of noise for quantum dot qubits. The most significant are nuclear spin noise and charge noise~\cite{kuhlmann_charge_2013, fernandez-fernandez_quantum_2022}. For tunnelling electrons, another noise contribution is electron-phonon scattering~\cite{hu_two-spin_2011, kornich_phonon-mediated_2014, he_theory_2023}. These noise sources particularly contribute dephasing noise, and lead to relatively short $T_2$ times compared to their relaxation times, $T_1$. The coherence times depend on the qubit encoding, with charge qubits having coherence times of $T_1 = 30~\textrm{ns}$ and $T_2= 7~\textrm{ns} $~\cite{chatterjee_semiconductor_2021}. For classical information as a low-dissipation classical bus, the dephasing noise is only crucial to maintain coherence for the perfect state transfer part of the protocol. Given a maximum tunnelling rate of $\hbar \Gamma = 360~\textrm{µeV}$, we find that the time for perfect state transfer, $T = (\pi /2) (\hbar /360 \textrm{µeV}) N \approx 3 N~\textrm{ps}$---see Section~\ref{sec:single_electron_state_transfer}. A chain of even 300 ions would still be significantly below the dephasing time of the charge qubits. However, the rate of voltage change required to freeze and unfreeze the chain would therefore be on the order of $\sim\textrm{ps}$, which is very fast~\cite{zou_simple_2017}. Reducing the tunnel coupling, and therefore the transfer time, is straightforward by increasing the confinement or increasing the distance between sites. The main source of error would actually be the inability to tune the tunnelling couplings accurately enough for perfect state transfer. The protocol could be applied sequentially, building up perfect state transfer chains such that the distances for each are significantly shorter than the noise that arises from mismatched tunnelling rates. The energetic cost would now be dependent on $N$, but with a very low prefactor. Even with perfect state transfer chains of only 10 quantum dots would provide an energetic advantage over shuttling. In the case of classical computing, we can disregard the phase information after each perfect state transfer step. 

Electrons can be confined in GaAs quantum dots for long times, on the order of seconds~\cite{camenzind_hyperfine-phonon_2018-2}. Hence classical bit flip errors are unlikely over the full length of the data bus. Repetition codes can be used to improve the fidelity of bit transmission. The protocol can be performed $m$ times and a majority vote of the outcomes can be used to determine the state.

\section{\label{sec:discussion}Discussion}
This work considers the energetic cost of state transfer protocols in quantum dot arrays. There are two clear and separate applications for these results. Firstly, to inform the design of quantum dot arrays for quantum computing---in particular, to minimise the on-chip dissipation (heat generation) which imposes demands on the cooling power of the refrigeration---and secondly, as a proposal for the limits of what is possible for energy-efficient data buses for classical information on semiconductor chips. 

The perfect state transfer protocols proposed give a theoretical energetic advantage to the current proposals of shuttling electrons. Recent work~\cite{boter_spiderweb_2022} on scalable quantum computing architectures in quantum dots considered the important issue of power consumption due to the control of a large number of quantum dots and the capacity to cool these devices. Low-dissipation data buses for transferring coherent quantum information would go some way to relaxing this constraint. 

Quantum dots and ion-trap chains have both recently been investigated as platforms for a potential energetic advantage in performing classical computations by using qubits and the coherent evolution of quantum systems~\cite{moutinho_quantum_2022,pratapsi_classical_2022}. Here, we consider the interconnects, another important and energetically costly component of a universal computer. Using the logical encoding of Ref.~\cite{moutinho_quantum_2022} for semiconductor quantum dots, we find that perfect state transfer offers a significant energetic scaling advantage compared to a classical data bus. The transfer of information via coherent quantum dynamics for classical data can also reduce a source of energetic overhead for using reversible quantum devices for classical computation. Without coherent quantum interconnects, the amount of data loading and unloading from classical information could be prohibitively expensive. On the other hand, if a reasonably-sized computational unit, such as an arithmetic-logic unit (ALU), could be implemented with reversible quantum dynamics and quantum coherent data buses, an energetic advantage becomes more attainable.

\end{document}